\documentclass[preprint,preprintnumbers,amsmath,amssymb]{revtex4}
\usepackage{graphicx}% Include figure files
\usepackage{dcolumn}% Align table columns on decimal point
\usepackage{bm}% bold math

\begin{document}
\title{Localized steady-state domain wall oscillators}
\author{J. He and S. Zhang}
\affiliation{Department of Physics and Astronomy, University of
Missouri-Columbia, Columbia, MO 65211}
\begin{abstract}
We predict a spatially localized magnetic domain wall oscillator upon
the application of an external magnetic field and a DC electric current.
The amplitude and frequency of the oscillator can be controlled by
the field and/or the current. The resulting oscillator could be used as an
effective microwave source for information storage application.
\end{abstract}
%\date{\today}
\maketitle

In a spin valve, a DC electric current generates a spin transfer
torque which can control or alter the magnetization dynamics of the
free layer. Above a critical current density, the spin transfer
torque is able to switch the direction of the magnetization,
generate spin wave excitations, and more interestingly, create a
steady-state precessional motion of the magnetization of the free
layer \cite{Kiselev}. For a ferromagnetic metal, the spin transfer
torque results in domain wall motion \cite{Yamaguchi,Parkin}, spin
wave excitations \cite{Ji,Li}, and wall transformation from one type
to another \cite{Klaui, He}. What is yet to demonstrate is whether a
{\em DC current} is able to create a {\em spatially localized}
domain wall oscillator. A well controlled and spatially localized
domain wall oscillator is very desirable for applications. For
example, the oscillatory magnetic field from the stray fields of a
localized domain wall oscillator can assist writing magnetic bits in
recording media. Here, we show that a stable and localized domain
wall oscillator is indeed possible by the combined applications of
the magnetic field and the current. We determine the relevant
parameters for this realization.

Domain wall motion driven by a magnetic field has been well studied
\cite{Schryer,Malozemoff,Nakatani,Beach}. When the magnetic field
exceeds a critical value, the domain wall overcomes the pinning
potential and begins to move to reduce the Zeeman energy. As a first
approximation, the domain wall moves uniformly and the velocity of
the wall is given by $\gamma H \Delta_0 /\alpha$ \cite{Schryer}
where $\gamma$ is the gyromagnetic ratio, $H$ is the magnetic field
along the wire, $\Delta_0$ is the wall width, and $\alpha$ is the
damping parameter. When the magnetic field is further increased
beyond the Walker-breakdown field, the wall motion is no more
uniform; instead, the wall velocity becomes oscillatory
\cite{Parkin}. However, the oscillation can not be sustained because
the wall motion or wall oscillation continuously decreases the
magnetic energy due to the presence of damping. After a typical time
scale of nanoseconds, the domain wall either stops precessing or
completely moves out of any finite regions, i.e., a {\em spatially
localized} wall oscillation driven by a DC magnetic field alone is
not realizable.

We show below that the spatially localized wall oscillator is
possible when a current is also applied. By properly choosing the
direction and magnitude of the current density for a given magnetic
field, the average velocity of the wall can be precisely controlled
at zero, and thus a stable and spatially localized oscillator can be
created. There is a key difference between the field-driven and
current-driven domain wall dynamics: the energy damped in a period
of wall oscillation can be compensated by the energy input from the
spin transfer torque, but no compensation occurs for the magnetic
field since the change of the Zeeman energy is zero for a full cycle
of oscillation.

To determine the relevant parameters for the creation of the
localized wall oscillator, we consider the dynamic equation of the
magnetization in the presence of field and current
\cite{Zhang,Thiaville,Bauer},
\begin{eqnarray}
\label{LLG} \frac{\partial {\bf M}}{\partial t}&=&-\gamma {\bf
M}\times {\bf H}_{eff}+\frac{\alpha}{\it{M}_{s}}{{\bf M}} \times
\frac{\partial {\bf M}}{\partial t}\\\nonumber
&&-\frac{b_{J}}{\it{M}^{\bf{2}}_{s}}{{\bf M}}\times\left({{\bf
M}}\times \frac{\partial{\bf M}}{\partial
x}\right)-\frac{c_{J}}{\it{M}_{s}}\bf{M}\times\frac{\partial
\bf{M}}{\partial \it{x}}
\end{eqnarray}
where ${\bf H}_{eff}$ is the effective magnetic field including the
external field, the anisotropy field, the magnetostatic field, and
the exchange field; $b_{J}=Pj_{e}\mu_{B}/eM_{s}$ and $c_{J}=\xi
b_{J}$, where $\emph{P}$ is the spin polarization of the current;
$j_{e}$ is the current density, $\mu_{B}$ is the Bohr magneton, $e$
is the electron charge, and $M_s$ is the saturation magnetization.
$\xi$ is a dimensionless constant which describes the degree of the
nonadiabaticity between the spin of the non-equilibrium conduction
electrons and local magnetization.

We solve above dynamic equation in two ways. First, we analyze a
simplified model based on the Walker's wall profile \cite{Schryer}:
this enables us to analytically determine the condition for the
formation of the wall oscillator. Numerical calculations are then
followed to verify our analytical results. In Walker's model, the
domain wall structure is characterized by two variables: the center
position of the wall $q(t)$ and the angle of the wall plane $\phi
(t)$. The wall width $\Delta_0$ is treated as a constant. With these
simplifications, one can write Eq.~(1) in terms of $q(t)$ and $\phi
(t)$,
\begin{eqnarray}
\Delta_0 \dot{\phi} + {\alpha} {\dot{q}} &=& \gamma \Delta_0 H - c_J
\\
\dot{q} - \alpha \Delta_0 \dot{\phi} &=& \gamma 2\pi M_s \Delta_0
\sin{2\phi} - b_J .
\end{eqnarray}
Eliminating $\dot{q}$ from above equations, we have
\begin{equation}
\dot{\phi} = \frac{\gamma}{(1+\alpha^2)} \left( H_0  - H_w \sin{2\phi}
\right)
\end{equation}
where we have defined field $H_{0} = H+(\alpha - \xi)b_J/(\gamma
\Delta_0)$, and the Walker breakdown field $H_w = \alpha 2\pi M_s$.
When $H_0 < H_w$, Eq.(4) has a steady state solution, i.e.,
$\dot{\phi}=0$ and $ \sin{2\phi}= H_{0}/H_w$; this is the solution
for the uniform motion of the wall with velocity: $\dot{q} = (\gamma
\Delta_0 H -c_J)/\alpha$. When $H_0 > H_w$, however, there is no
steady-state solution; this is known as the Walker breakdown
\cite{Schryer}. The direct integration of Eq.~(4) yields
\begin{equation}
\tan\phi(t) = C_1 + \sqrt{1- C_1^2} \tan(\omega t)
\end{equation}
where $C_1= H_w/H_0$ and $\omega = \gamma \sqrt{H_0^2 - H_w^2}
/(1+\alpha^2)$. Equation (5) indicates that the angle $\phi$
increases a $\pi$ in one period $T_{\phi} =\pi/\omega$. Thus the
average angular velocity is
\begin{equation}
\dot{\bar{\phi}} = \pi/T_{\phi} = \frac{\gamma}{(1+\alpha^2)}
\sqrt{H_0^2 - H_w^2}.
\end{equation}
By averaging Eq.~(2) over one oscillation period and by using
Eq.~(6), we arrive at the condition for the localized oscillator
(setting $\dot{\bar{q}}=0$),
\begin{equation}
b_{J} = \frac{\gamma \Delta_0}{\xi} \left( H - \frac{\sqrt{H_0^2 -
H_w^2}}{1+\alpha^2} \right).
\end{equation}
The amplitude of the oscillation can be obtained by solving for $q$
when $\dot{q}=0$. There are two solutions; their difference (divided
by 2) is identified as the oscillation amplitude $q_{amp}$. From
Eqs.(2), (3) and (7), we have
\begin{equation}
q_{amp} \simeq \frac{\Delta_0 H_w}{2\alpha H}.
\end{equation}

To further gain the insight on the solution of the localized wall
oscillation, let us consider the change of wall energy in one period
of oscillation. Multiply Eq~(2) by $\dot{q}$ and multiply Eq.~(3) by
$\Delta_0 \dot{\phi}$, and then substrate the resulting equations
from each other, we have
\begin{equation}
\pi \Delta_0 b_J = \alpha\int_0^{T_{\phi}} dt ( \dot q ^2 +
\Delta_0^2 \dot \phi^2).
\end{equation}
The right hand side represents the domain wall energy loss in one
cycle, which has to be compensated by the work done by the spin
torque on the left hand side. Note that the external field does not
contribute any work in a complete cycle; this is the physical reason
why the field alone is unable to sustain a localized domain wall
oscillation.

Up till now, we have considered the solution of localized wall
oscillations in an ideally uniform film or wire. The center position
of the wall oscillation is arbitrary as long as Eq.~(7) is met. Any
spatial variation of the parameters would lead to a drift of wall
position. In order to stabilize the center of the wall oscillation
at a desired location, one needs to design a structure that can
suppress the drifting of the wall center but maintain the wall
oscillation. At first, one might consider a local pinning to trap
the oscillator, for example, by using a higher anisotropy material
in a small region. However, we find that such local pinnings are not
effective at all. If the pinning is strong, the wall oscillation is
completely destroyed and a static domain wall will be formed at the
pinning site. If the wall oscillation persists over a weaker pinning
potential, the wall center remains unstable against a small
fluctuation of parameters. The reason is that the amplitude of the
wall oscillation is several times larger than the wall width, see
Eq.~(8), so that the local pinning does not affect the oscillation
significantly. We thus propose a scheme to stabilize the oscillation
by designing a spatially varying damping parameter--this can be
achieved via gradient doping of rare-earth impurities in
ferromagnets \cite{Reidy}. We argue below that the wall oscillation
is spatially stable in this design.

Consider a spatial dependence of the damping parameter $\alpha (q)$.
For a fixed current $b_J$ and field $H$, the center of the
oscillation will be located at a certain position $q=x_0$ so that
Eq.~(7) is satisfied when $\alpha = \alpha (x_0)$. Then if there is
a fluctuation, for example, the current density is slightly
increased, the wall center will move along the direction of electron
current. The equation of motion for the center of the wall is, from
Eqs.~(2) and (3),
\begin{equation}
\ddot {\bar{q}} \simeq \gamma\Delta_0 H_{0}\left[2+\frac{(2\pi
M_s)^2}{H_0^2}\right]\frac{d\alpha}{d\bar{q}}\dot {\bar{q}}
\end{equation}
where $ H_w\ll H_0$ and $\alpha\ll 1$ have been used. Clearly, if
$d\alpha/d\bar{q}<0$, i.e., the damping parameter increases along
the direction of the electron current (note $d\bar{q}<0$ if $b_J$ is
increased), the drifting velocity of the center of oscillation
$\dot{\bar{q}}$ exponentially decays to zero. The wall oscillates
around a new position near the original oscillation center, where
Eq.~(9) is still satisfied on average.

Next we numerically solve Eq.~(1) to validate our analytical
results. We choose a magnetic wire whose width and thickness are
sufficiently small so that the transverse wall is energetically
favorable compared to the vortex wall. We also choose $\alpha = \xi
= 0.02$ throughout the simulation; when $\alpha \neq \xi$, there is
no qualitative difference on the behavior of the wall motion except
that the effective magnetic field has an additional term given by
$(\alpha - \xi)b_J/(\gamma \Delta_0)$. In Fig.~1, we show the
typical wall velocity $\dot{\bar q}$ as a function of the magnetic
field with and without the current. The linearly increasing of
$\dot{\bar q}$  at small fields represents the uniform steady-state
motion of the wall. When $H_0
>H_w$, the average velocity decreases since the domain wall does a
reciprocative motion, i.e., the wall motion is oscillatory but
$\dot{\bar q}$ is generally non-zero -- the oscillator is not
localized. At certain values of the field and current, we find
$\dot{\bar q}=0$, indicated by the mark ``$\it X$'' in Fig.~1. In
the insert, we show the oscillation of the wall center position
$q(t)$ around a fixed point ($x=0$). We notice that the current
density required for the localized oscillator is relatively small,
$b_J=20 m/s$ or $j_e=3.9\times 10^7 A/cm^2$ (for $P=0.7$), compared
to the experiments on current-driven domain wall motion
\cite{Parkin,Yamaguchi,Hayashi,Tsoi2}; this makes experiments of
searching for the localized oscillator easily accessible.

We describe the stability of the localized oscillator by choosing a
linearly varying damping parameter $\alpha (x) = 0.02\times(1-
x/L)$, where $L$ is the wire length and $-L/2 < x < L/2$. We first
consider the field $H$ and the current $b_J$ such that the
oscillator is localized at $x=0$, see the point ``$\it X$'' in
Fig.~1. When we slightly vary the current density or the magnetic
field, the oscillator will relocate to a new center position $x_1$
near $x=0$. The time evolution of the displacement $q(t)$ at this
new position (not shown) is similar as the insets of Fig.~(1). We
show in Fig.~2 the position of the wall oscillation center,
amplitude and frequency of the localized oscillator for the varying
magnetic fields and currents. These results are in good agreements
with the analytical results, Eqs.~(6), (7), and (8).

Finally, we emphasize that the localized domain wall oscillators
proposed here is quite different from the previous work
\cite{Tatara1,Tatara2} where either an AC current or an AC magnetic
field is used as a driving force. In those cases, a strong pinning
potential via geometrical confinement is used to localize the domain
wall and the oscillation of the wall is simply a response to the
oscillatory external force (AC fields or currents). Our proposal
here is to generate the wall oscillation by a DC magnetic field and
to localize the oscillator by a DC electrical current. A spatial
varying damping parameter can effectively stabilize the oscillator.

This work is partially supported by DOE and Seagate Technologies Inc.

\pagebreak

\pagebreak
\bigskip
\noindent {Figure Caption}

\bigskip
\noindent{FIG.1} (Color online) Average velocity of the domain wall
$v=\dot{\bar q}$. The parameters are: $M_s=800$ (emu/cc), $H_K=500$
(Oe), $A=1.3\times 10^{-6} $ (erg/cm). Note we have chosen a reduced
Breakdown field similar to the experimental value $H_w=9$ (Oe)
\cite{Hayashi}. The fitting curves are the analytical solutions of
$\dot {\bar{q}}$ from Eqs.~(2) and (6), and the fitted wall width
$\Delta_0=20$nm. Insets show the time evolution of the displacement
$q(t)$ at the point ``$\it X$''.
\bigskip

\bigskip
\noindent{FIG.2} (Color online) Position of the wall oscillation
center $x_1$, amplitude $q_{amp}$ and frequency $f=\omega/\pi$ of
the oscillator as a function of the field and current: (a), (b) and
(c) are for a fixed current, and (d), (e) and (f) for a fixed field.
Damping parameter is $\alpha (x) = 0.02\times(1- x/L)$, where $L =
0.8 \mu$m.

\bigskip
\newpage

\begin{figure}
\centering
\includegraphics[width=14cm]{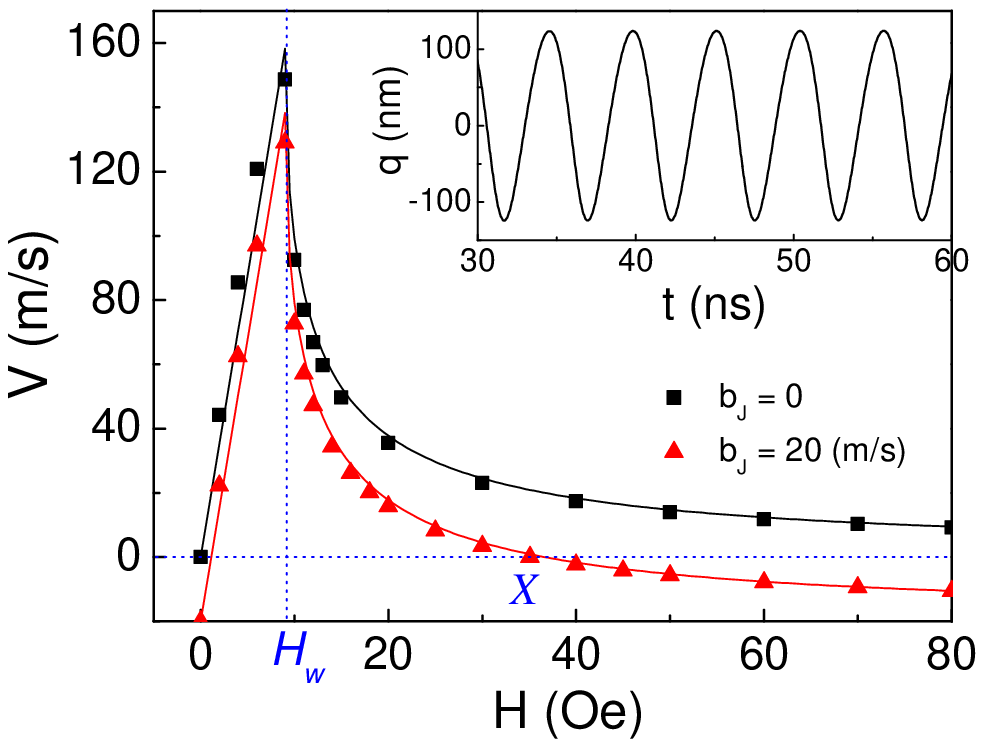}
\caption{}
\end{figure}
\bigskip
\bigskip
\pagebreak
\newpage
\newpage
\begin{figure}
\centering
\includegraphics[width=14cm]{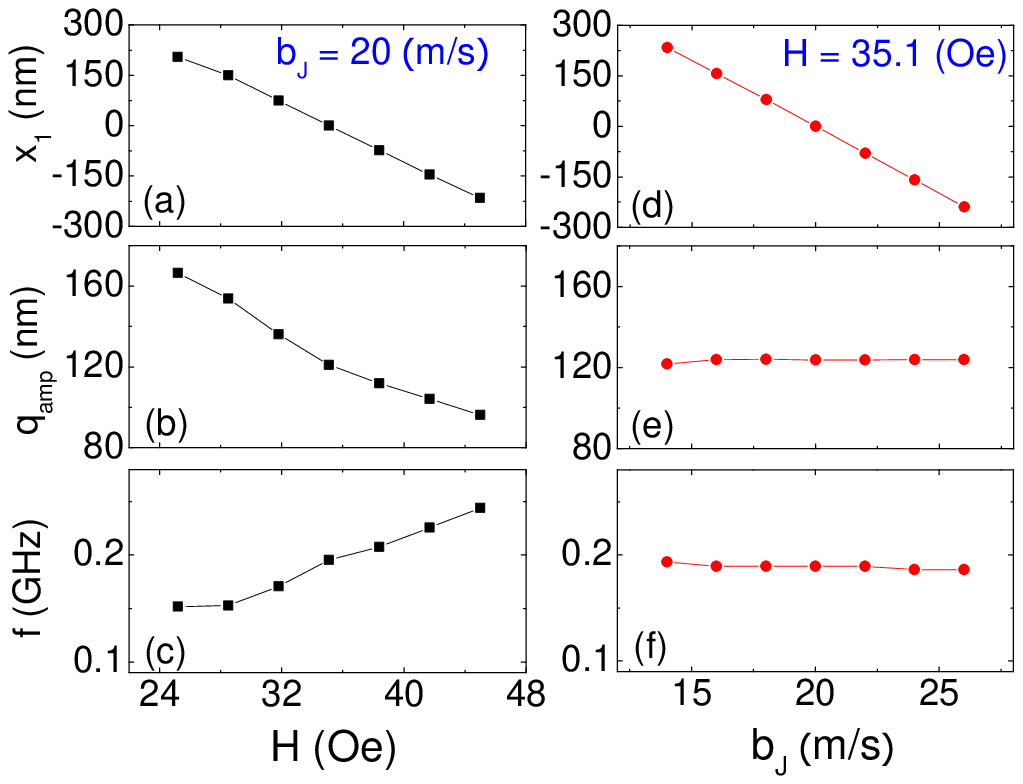}
\caption{}
\end{figure}

\end{document}